\documentclass[5p]{elsarticle}
\usepackage{amsmath,amssymb,mathrsfs,graphicx,epstopdf,placeins,float}
\usepackage{natbib}
\usepackage[pdfstartview=FitH]{hyperref}

 \newcommand{\beq}[1]{\begin{equation}\label{#1}}
 \newcommand{\eeq}{\end{equation}}
 \newcommand{\bea}[1]{\begin{eqnarray}\label{#1}}
 \newcommand{\eea}{\end{eqnarray}}
 \newcommand\figcaption{\def\@captype{figure}\caption}
 \newcommand\tabcaption{\def\@captype{table}\caption}

\begin{document}
\title{A Semi-analytical Solution to Classic Yang-Mills Equations with Both Asymptotical Freedom and Confining Features}

\date{\today}
\author{Ding-fang Zeng\footnote{dfzeng@bjut.edu.cn}}
\address{Physics Department, Beijing University of Technology}
\begin{abstract}
 It is well known that confinings and asymptotic freedom are properties of quantum chromo-dynamics (QCD). But hints of these features can also be observed at purely classic levels. For this purpose we need to find  solutions to the colorly-sourceful Yang-Mills equations with both confining and asymptotic freedom features. We provide such a solution in this paper which at the near-source region is of serial form, while at the far-away region is approximately expressed through simple elementary functions. From the solution, we derive out a classically non-perturbative beta function describing the running of effective coupling constant, which is linear in the couplings both in the infrared and ultraviolet region.
 \end{abstract}
 \maketitle
 \allowdisplaybreaks
 
\section{Introduction}

As is well know, exact solutions to the Einstein equation shape our knowledge structure of general relativities remarkably. Very naturally, we expect that if exact solutions to the classic Yang-Mills equation like those of Schwarzschild or Riessner-Nordstr\"om to Einstein equations could be found, they will also be very meaningful for our understanding of strong interactions. However, in constrasting with the richfulness of exact solutions to Einstein equations, exact solutions to the classic Yang-Mills equation are very few. The old papers include those on the construction of colorless-monopoles \cite{SU2monopole}-\cite{PSsolution}, the colorful black-holes \cite{sphericalSymmetricYMfield}-\cite{particleLikeSolution}, and the non-gravity but horizon-carrying Yang-Mills-Scalar black holes \cite{Yang-Mills-BH}. The relatively new papers include \cite{gaussLawYM}, \cite{yangMillsCurvedSpace}, \cite{solByMathematicians1} and \cite{solByMathematicians2} et al. References \cite{gaussLawYM} and \cite{yangMillsCurvedSpace} emphasizes the physical aspects, while \cite{solByMathematicians1} and \cite{solByMathematicians2} put the main focus on mathematics. The purpose of this paper is to provide solutions to the purely classic Yang-Mills theories with both confining and asymptotical freedom features.

A very important motivation of this work is reference \cite{classicConfinement} (see also \cite{EFTideal} for earlier ideas in a totally different environment), in which G. Dvali et al observed that  by studying the classic field stimulated by fixed external sources, many features of the underlying quantum theories, such as the renormalization group structure, the phenomenon of dimensional transmutation, running coupling constant, asymptotic freedom and classic et al can all be exhibited out at the purely classic level. G. Dvali et al take the $\lambda\phi^4(\lambda<0)$ theory as examples to illustrate these ideals. Obviously, it will be more instructive if for Yang-Mills theories we can also see asymptotical freedom and confinement et al at purely classic levels. 

The organization of this paper is as follows. This section explains our motivation for this work, the next section introduce the basic ideal of Dvali et al's work on observing confinings at purely classic levels. Section III provides the basic action and classic equation of motions (spherical symmetric ansatz) for Yang-Mills fields coupled with external quarks. Section IV provides three exact solutions to the classic equations, the third one of with has both confining and asymptotical features. Section V provides a trial ideal to get classical beta functions describing the running of couplings of the theory. Section VI is our main conclusions. The final is an appendix devoted to prove the confinings of pure Yang-Mills theories at the purely classic levels.

\section{Dvali et al's ideal of classic running couplings and confinings}
Generally, it is thought that, confinings, asymptotical freedoms, running coupling constants and dimensional transmutations are all quantum effects, they could be observed only at quantum levels. However, in reference \cite{classicConfinement}, Dvali et al proposed that these phenomenaes can also be observed at purely classic levels. For this purpose, Dvali et al considered a classic $\lambda\phi^4$ theory with $\lambda<0$,
 \beq{}
 S=\int\!\!d^4\!x\big(\frac{1}{2}\partial_\mu\phi\partial^\mu\phi-\frac{1}{4}\lambda_0\phi^4
 +4\pi Q\phi\big)
 \eeq
 where $Q$ is the charge of an external point source, when $Q\gg1$, the classic part of the field stimulated by $Q$ will dominate over its quantum fluctuation. Consider the spherical symmetric ansatz, the radial profile of the field will be determined by
 \beq{}
 \frac{1}{r^2}\frac{d}{dr}(r^2\frac{d\phi}{dr})=-Q\frac{\delta{(}r{)}}{r^2}+\lambda_0\phi^3
 \label{phi4Eq}
 \eeq
 If $\lambda_0Q^2\ll1$, then we can use iteration method to solve this equation. The result reads
 \beq{}
 \phi=\frac{Qf{(}r{)}}{r}\equiv\frac{Q_\mathrm{eff}}{r}
 \label{QeffDefinition}
 \eeq
 \beq{}
 f(r)=1+\alpha_0r\int_r^\infty\!\!\Big[\int_{r_0}^{r'}\frac{f(r'')^3}{r''}dr''\Big]\frac{dr'}{r'^2}
 -C(\alpha_0)
 \label{integralEqf}
 \eeq
with $r_0$ being an arbitrarily chosen length scale and the form of $C(\alpha_0)$ fixed by the condition $f(r_0)=1$. Obviously, if $\lambda_0=0$, then $f\equiv1$ and $C(\alpha_0)\equiv1$, while $\phi$ reduces to the simple columb potentials. If $\lambda_0<0$, then $\alpha_0\equiv\lambda_0Q^2<0$, $f<1$. In physical languages, in a free scalar field theory, the charge of external point sources is a constant. While a scalar field theory with nonlinear self-interactions, the charge of the same external point source depends on the scale at which it is measured. 

Defining an effective coupling constant
\beq{}
\alpha_0f^2{(}r{)}\equiv\alpha_\mathrm{eff}{(}r{)}
\eeq
In weak coupling conditions, from the iteration formula \eqref{integralEqf}, Dvali et al derive out that
\beq{}
\alpha_\mathrm{eff}{(}r{)}=\sum_0^\infty\alpha^{n+1}(r_0)g_n(x)
,~x\equiv\ln\frac{r}{r_0}
\label{alphEffDefinition}
\eeq
\beq{}
g_0=1
,~
g_1=2
,~
g_2=4x^2+6x
\eeq
\beq{}
g_3=8x^3+30x^2+48x
,\cdots
\eeq
Very coincidently, these coefficients satisfy the the renormalization group structure equaion
\beq{}
\frac{dg_{n+1}(x)}{dx}=\sum_{k=0}^n
(k+1)g'_{n+1-k}(0)g_k(x)
\label{rgsEq}
\eeq
This is a set of highly nontrivial constraints on $\{g_n\}$ and a necessary condition for the dependence of $\alpha_\mathrm{eff}$ on $r$ be interpreted as the scale dependence of renormalization groups. Approximating to ``one loop'' order --- one time of iterations --- equation \eqref{rgsEq} has solution
\beq{}
\alpha{(}r{)}=\frac{\alpha_0}{1-2\alpha_0x}
=\frac{-\lambda_0Q^2}{1+2\lambda_0Q^2\ln(r/r_0)}
\eeq
Obviously, this is almost the same as the running coupling constants of QCD. In the case $\lambda_0<0$, asymptotical freedo corresponds to the fact that $\alpha{(}r{)}\rightarrow0$ as $r$ becomes smaller and smaller. 

While in the strong coupling region, according to the classic equation of motion \eqref{phi4Eq} and definitions \eqref{QeffDefinition}, \eqref{alphEffDefinition}, it can be proved that
\beq{}
\alpha''-\frac{\alpha^{_\prime2}}{2\alpha}-\alpha'+2\alpha^2=0
,~\alpha\equiv\alpha(\ln\frac{r}{r_0})
\eeq
which after defining that $\beta\equiv\frac{d\alpha}{d\ln(r/r_0)}$, can just be looked as the classical renormalzation group equation for the coupling constant $\alpha$,
\beq{}
\beta=2\alpha^2+\frac{1}{2}\Big[\frac{d(\beta\cdot\beta)}{d\alpha}-\frac{\beta\cdot\beta}{\alpha}\Big]
\eeq
From this exact renormalization group equation, Dvali et als find out that in the strong coupling region
\beq{}
\alpha{(}r{)}\simeq\mathcal{O}(1)\big(\frac{r}{r_c}\big)^\frac{2}{3}\cos^2\big(\frac{r}{r_c}\big)^\frac{2}{3}
\eeq
Basing on this result, Dvali et als find that the total energy of an isolated point-like external source diverges as $r^\frac{1}{3}_{r\rightarrow\infty}$. While by some qulitative arguments, they conclude that the interaction energy of a ``quark-antiquark''  seperated by $\ell$ distances in this $\lambda\phi^4$ theory should behaves as $\ell^\frac{1}{3}$. Using knowledges from this two strong coupling considerations, Dvali et al stated that at least, hints for confinings of this theory has been illustrated at the purely classic levels. They also make some comments on this ideal to QCD theories and point out that, due to the interaction structure difference of QCD from those of single-component scalar field theories, the interaction energy of a quark-antiquark pair need not behaves as $\ell^\frac{1}{3}$.

What we will do in the following is to apply the basic ideal of Dvali et als to the Yang-Mills fields coupled with external quarks. But due to the complexities of classic Yang-Mills equation relative to those of the scalar field theory, our calculations cannot be carried out totally parallel with those of Dvali et als. For example, (i) we find no iteration method similar to those of Dvali et als, but have to use a direct serial expansion method to solve classic Yang-Mills equations. (ii) due the lack of iteration method, we do not know how to use arguments like those of Dval et als to infer the form of quark-antiquark interaction energies. 

\section{Equations of motion and the solution ansatz}

Let us begin our non-abelian stories from the basic action of classic Yang-Mills fields coupled with external quarks
\beq{}
\mathcal{L}=-\frac{1}{2g^2_{_{Y\!M}}}\mathrm{tr}F_{\mu\nu}F^{\mu\nu}
+\frac{2}{g_{_{Y\!M}}}\mathrm{tr}A_\mu J^{\mu}
\eeq
where $A_\mu\equiv A^a_\mu t^a$,~$J_\mu=J^a_\mu t^a$,
~$J^{a\mu}=\bar\psi_i\gamma^\mu t^a_{ij}\psi_j$, with $t^a$s denoting the generator of $SU(N)$ group while
\beq{}
F_{\mu\nu}\equiv\partial_\mu A_\nu-\partial_\nu A_\mu
+i[A_\mu, A_\nu]
\eeq
Using these notations, the covariant Yang-Mills equation can be written as
\beq{}
D_\mu F^{\mu\nu}\equiv\frac{1}{\sqrt{-g}}\partial_\mu(\sqrt{-g}F^{\mu\nu})+i[A_\mu,F^{\mu\nu}]=g_{_{YM}}J^\nu
\label{covYMeq}
\eeq
where $\sqrt{-g}$ is the root of metric determinant in some specific coordinating system.

For our purpose, we need to find solutions to the above Yang-Mills equation with point like static quark sources. The corresponding 4-component quark-current vector can be written as
\beq{}
J^{\mu}=(Q^at^a4\pi\delta(\vec{\mathbf{r}}),0,0,0)
\eeq
By symmetry reasoning, we know that gauge fields stimulated by this quark should be spherically symmetric. According to reference \cite{sphericalSymmetricYMfield} and \cite{particleLikeSolution}, the most simple color-fields with this symmetry could be written as
\beq{}
A^a_\mu t^a dx^\mu=A_tdt+A_rdr+A_\theta d\theta+A_\phi d\phi
\eeq
\beq{}
A_t=D\cdot\frac{J(r)}{2r}
~,~~A_r\equiv0
~,~~A_\theta=-\frac{i}{2}(C-C^\dagger)K(r)
\label{A0Ansatz}
\eeq
\beq{}
A_\phi=-\frac{1}{2}\big[(C+C^\dagger)K(r)\sin\theta+D\cos\theta\big]
\label{AphiAnsatz}
\eeq
where $J( r )$ and $K( r )$ are two functions to be determined latter, they has no relevance with the usual Bessel Function. We will fix our gauge by requiring that when the quarks points to a pure electric-direction in the color-space, the relevant color field reduces to those of a pure point electric-charges.

By properly choosing the direction of source quarks' charge $Q^a$ in the color-space, we can always write the matrices $C$ and $D$ involved in the above expressions as follows (noting we use $\bar{i}$ denoting $N\!-i$)
\beq{}
D\equiv\mathrm{diag.}\{N-1,N-3,\cdots,3-N,1-N\}
\eeq
\beq{}
C\equiv\left[\begin{array}{ccccc}
0&\!\sqrt{1\!\cdot\!\bar1}\!&~&~&~\\~&0&\!\sqrt{2\!\cdot\!\bar2}\!&~&~\\
~&~&\!\ddots\!&\!\ddots\!&~\\~&~&~&0&\!\sqrt{\bar{1}\!\cdot\!1}\\
~&~&~&~&0
\end{array}\right]_{\!N\times N}
\eeq
\if0\beq{}
D\equiv\left[\begin{array}{lllll}\!N\!-\!1\!&~&~&~&~\\
~&\!N\!-\!3\!&~&~&~\\~&~&\ddots&~&~\\
~&~&~&\!3\!-\!N\!&~\\~&~&~&~&\!1\!-\!N\!
\end{array}\right]_{\!N\times N}
\eeq
\fi
substituting this ansatz into the covariant Yang-Mills equation, we can easily derive out equations satisfied by $J(r)$ and $K(r)$
\begin{subequations}\bea{}
&&\hspace{-5mm}r^2J''-2JK^2=g_{_{YM}}Qr\delta(r)
\label{eq1Quark}\\
&&\hspace{-6mm}r^2K''-K(K^2-1-J^2)=0
\label{eq2Quark}
\eea\label{eqQuark}
\end{subequations}
where $Q$ is now only a number quantifying the amount of color-charges, its direction in the color-space has been specified when we choose the form of matrices $C$ and $D$.
Although this is a sourceul equation,  we will neglect the source term when solving differential equations. This is very similar to the case of solving Poinsson equations $\nabla^2\phi=eQr\delta(r)$ to get Column potentials in electrostatics or solving Einstein equations $R_{\mu\nu}-\frac{1}{2}g_{\mu\nu}R=0$ to get Schwarzschild metric in general relativity. The difference is, in Column potentials and Schwarzschild metric, we have definite scheme to relate the solution parameter with the source chages (masses is looked as charges of gravity). While in the Yang-Mills case, due to the feature of asymptotical freedom and confinings, we still have no appropriate schemes to relate the solution parameter $J_2$, $K_2$ in the following to the source charge $Q$.

\section{Spherical symmetric solutions and hints for confinings at the purely classic levels}
Now let us give out the first two simple but non-trivial solution
\bea{}
(i)~J\equiv0,~K\equiv0,~~~~~~~~~~
\label{solQuarkColumn}
\\
(ii)~J=ar+b,~K=0~(b\neq0)
\label{solDyonColumn}
\eea
Their non-triviality can be looked out from their corresponding field strength
\bea{}
(i)F^a_{0i}t^a\equiv0,~F^{a}_{\theta\phi}t^a=\frac{1}{2}D\sin\theta
\label{fStrengthQuarkColumn}~~~~~~
\\
(ii)F^a_{0r}t^a=\frac{D}{2}\frac{b}{r^2}
,~F^a_{\theta\phi}t^a=\frac{1}{2}D\sin\theta
\label{fStrengthDyonColumn}
\eea
The first solution describes objects with constant color-magnetic field, while the second one describes  object with both constant color-magnetic field and Column-like color-electric field. Obviously, the solution parameter $b$ is equal to the source charge $Q$. If we calculate energies stored in this two field configurations (see expression \eqref{energyCalculation1} in the following), we will see that both of them contain singularity at $r\rightarrow0$, i.e. $E\propto\int_0^\infty\!dr/r^{2}$. This is a singularity similar to that of point-like electric charges in classic eletrodynamics. 
In reference \cite{yangMillsCurvedSpace}, the solution (ii) is thought to be confinings, but just as we can see from the field strength expression, this is not the case.

\begin{figure}[tb]
\parbox{0.22\textwidth}{\includegraphics[scale=0.77]{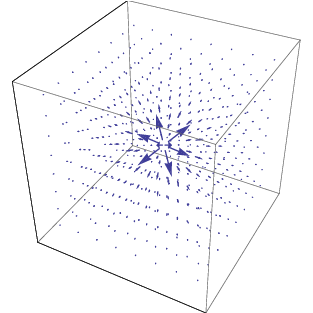}}
\parbox{0.22\textwidth}{\includegraphics[scale=0.77]{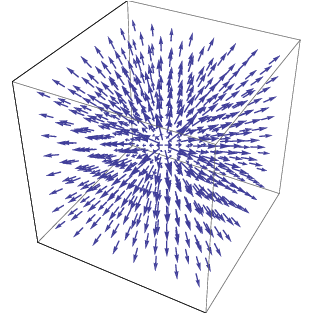}}
\parbox{0.5\textwidth}{
\includegraphics[scale=0.77]{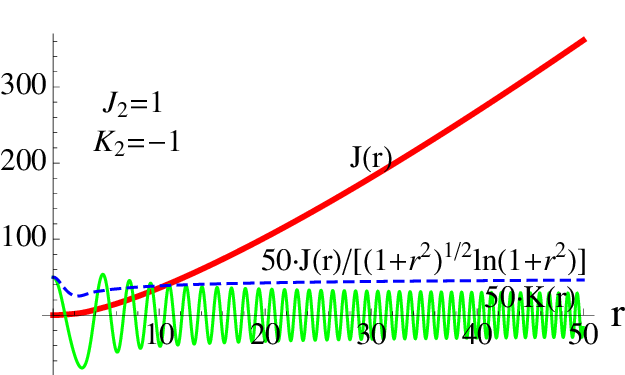}
}\caption{(Color online)Upper part, the left panel is the field strength of the usual Columb field, the right panel is that of color-electric field with running, asymptotical freedom coupling constant. Downer part is a typical numerical solution to the classic Yang-Mills eqs\eqref{eqQuark} with the boundary conditions setting as $J(r)\xrightarrow{r\rightarrow0}J_2r^2$, $K(r)\xrightarrow{r\rightarrow0}1+K_2r^2$. The dashed line in the figure displays a reference function which is approximately $\propto\frac{J(r)}{r\ln(r)}\rightarrow$ constant. Changing the value of $J_2$, $K_2$ does not lead to the change of $J(r)$'s asymptotical behavior and $K(r)$'s oscillating feature. Flipping the sign of $J_2$ flips $J(r)$'s sign correspondingly. }
\label{figNsol}
\end{figure}

The previous two solutions are both non-confining. Maybe the more attractive solutions should be the following two-parameters solution family with both asymptotical freedom and confining features
\begin{subequations}
\bea{}
&\hspace{-14mm}(iii)J(r)\stackrel{r\rightarrow0}{=}\sum\limits_{n=2,4}^{6,\cdots} J_nr^n
,~
K(r)\stackrel{r\rightarrow0}{=}\sum\limits_{n=0,2}^{4,\cdots} K_nr^n
\label{rZeroBehavior}
\\
&\hspace{-5mm}\Big\{\begin{array}{l}
J^2(r)\!\stackrel{r\rightarrow\infty}{=}\!
\frac{\cos^2[A]}{A'}-\!1\!+\!r^2\!\Big[\!A'^2\!+\!\frac{1}{2}\frac{A'''}{A'}\!-\!\frac{3}{4}\big(\frac{A''}{A'}\big)^{\!2}\!\Big]
\\
K(r)\stackrel{r\rightarrow\infty}{=}\frac{\cos[A]}{\sqrt{A'}}
,~A\stackrel{r\rightarrow\infty}{\approx} r\ln(r)
\label{rInfBehavior}
\end{array}
\eea
\label{confiningSolution1}
\end{subequations}
With the corresponding field strength given by
\beq{}
(iii)F_{0r}=-\frac{rJ'-J}{2r^2}D
,~F_{0\theta}=\frac{JK}{2r}(C+C^\dagger)
\label{fStrength}
\eeq
\beq{}
F_{0\phi}=-i\frac{JK}{2r}(C-C^\dagger)\sin\theta
,~
F_{r\theta}=-i\frac{K'}{2}(C-C^\dagger)
\nonumber
\eeq
\beq{}
F_{r\phi}=-i\frac{K'}{2}(C+C^\dagger)\sin\theta
,~F_{\theta\phi}=-\frac{(K^2-1)}{2}D\sin\theta
\nonumber
\eeq
Although converges only in a small region around $r=0$, the serial expressions \eqref{rZeroBehavior} exactly satisfy equations \eqref{eq1Quark} - \eqref{eq2Quark} as long as
\bea{}
J_4=\frac{2}{5}J_2K_2,~
J_6=\frac{6J_2K_2^2}{35}-\frac{J_2^3}{70}
\\
\nonumber
J8=\frac{104J_2K_2^3}{1575}-\frac{4J_2^3K_2}{225}
,\cdots
\eea
\bea{}
K_0=\pm1,~K_4=\frac{3K_2^2-J_2^2}{10}
,~
K_6=\frac{K_2^3}{10}-\frac{3J_2^2K_2}{35}
\\
K_8=\frac{37J_2^4-606J_2^2K_2^2+413K_2^4}{12600}
,\cdots
\nonumber
\eea
with $J_2$ and $K_2$ being two parameters characterizing the color-electric and color-magnetic charges of the source. The analytical expressions \eqref{rInfBehavior} exactly satisfies eq\eqref{eq2Quark} but satisfies \eqref{eq1Quark} only up to an order $r\cos[r\ln(r)]$ oscillating errors. It can be easily proved that $J(r)$ in this expression approaches $r\ln r$ asymptotically.  FIG \ref{figNsol} displays one of our numerical solutions explicitly.

Since we have fixed our gauges by requiring that as the quark-source point to a purely electric-direction in the color space, the relevant field strength reduces to the usual columbs one. We can read out the asymptotics of the Yang-Mills theories from eq\eqref{fStrength} and the serial solution \eqref{rZeroBehavior}. For example, from the former equations we see that both the color-electric and color-magnetic field have behavior $F_{0r}$, $\frac{F_{\theta\phi}}{r^2}\xrightarrow{r\rightarrow0}{}$ constant instead of the Column-like increasing $r^{-2}$. This is a feature of asymptotical freedoms. Another way to see this is using the fact that $J(r)\xrightarrow{r\rightarrow0}r^2$ and that $rJ'-J\xrightarrow{r\rightarrow0}0$ has the  meaning of effective color-electric charges. The color-magnetic charge has similar properties.The confinement of the solution can be observed from the corresponding field-configuration energies, which are defined and can be calculated \cite{PSsolution} as follows
\beq{}
E\equiv\int\!\!d^3\!\mathbf{x}\,T^{00},~
T^{\mu\nu}=\frac{2}{\sqrt{-g}}\frac{\delta(\sqrt{-g}\mathcal{L})}{\delta g_{\mu\nu}}
\label{energyCalculation0}
\eeq
\bea{}
E&\!\!=\!\!\!&\int_0^\infty\!\!\!\!\!dr\frac{(rJ'\!\!-\!J)^2\!+\!2J^2\!K^2
\!+\!2r^2\!K'^2\!+\!(K^2\!-\!1)^2}{r^2}
\label{energyCalculation1}
\\
&\!\!\approx\!\!&\int_0^\infty\!\!\!\frac{dr}{r^2}\Big[r^2+2r^2\ln(r)+2r^2+1\Big]
\label{energyCalculation2}
\eea
In the last approximation, we used the fact that $J(r)\xrightarrow{r\rightarrow\infty}r\ln(r)$, $K(r)\rightarrow\cos[J(r)]/\sqrt{J'(r)}$ and indicated contributions only around $r\rightarrow\infty$. Obviously, this energy is infrared divergent. Since it is the energy of chromo-fields stimulated by colored objects, its diverging means that such objects cannot exist as isolating objects, and thus confinings, or at least the hints for confinings at the purely classic levels. It is worth noting that the energy defined by equation \eqref{energyCalculation0} is gauge-independent, so the confinement claiming from its divergences is also a gauge-independent declaration.

At quantum levels, the confining feature of QCD theories is established through the provements of area laws for the expectation value of wilson loops. Physically this is equivalent to prove the linear law for interaction potentials of a quark-antiquark pair and the formation of flux tubes. However, to illustrate this phenomena at the purely classic levels, we have to solve the Yang-Mills equation with a double-center source
\beq{}
D_\mu F^{\mu\nu}=g_\mathrm{YM}Q^at^a4\pi\{\delta(\vec{\ell}/2)-\delta(-\vec{\ell}/2),0,0,0\}
\eeq
However, since in this case the solution is no longer be spherical symmetric, our analysis in this and previous sections is no longer useful. And since we find no iteration method similar those of Dvali et als, we cannot use arguments similar to them to get the $E_{q\bar{q}}\propto\ell^\alpha$ conclusions. The iteration method allows them to substitute the zeroth order solution, i.e. the dipole potential function $\phi\propto r^{-2}$ into the iteration fomula and get the asymptotic form of the scalar ``quark-antiqurk'' potential, $E_{q\bar{q}}\propto\ell^\frac{1}{3}$. We delay this task of developing iteration method for future works.

\section{Running couplings and classical beta function}

Now, comes our plaguy question, how to relate the solution parameter $J_2$, $K_2$ with the source charge $Q$? Simply using Gauss' law and flux integration does not work here. Because, due to the running of effective couplings, fluxes on various surface enclosing the source charge are not constants. While $g_{_{YM}}$ and $Q$ always appear as a whole, see eq\eqref{eqQuark}. However, this running also suggests us using renormalization group method to relate $J_2$, $K_2$ and $Q$. To use this method \cite{classicConfinement}, we need first define appropriate effective coupling constants(in terms of $J(r)$, $K(r)$ functions) and prove that their runnings have renormalization group structures (RGS). As a try, we define
\beq{}
(rJ'-J)_{r=r_0}=Q,~(K^2-1)_{r=r_0}=zQ
\label{gem0definition}
\eeq
\beq{}
r J'-J\equiv g_e(r)Q,~
K^2-1\equiv g_m(r)zQ
\label{gemEffective}
\eeq
with $z$ being a parameter to be determined latter. Obviously, eq\eqref{gem0definition} implements the goal of connecting the source charge $Q$ to the solution parameter $J_2$ and $K_2$  While $g_{e}(r)$ and $g_{m}(r)$ defines the effective running color-electric or color-magnetic couplings.  Reversing eqs\eqref{gem0definition} we can write $J_2r_0^2=\sum c^z_{jn}$\!$Q^n$, $K_2r_0^2$ $=\sum c^z_{kn}$\!$Q^n$. Substituting them into eq\eqref{gemEffective} we will get
\bea{}
&&\hspace{-5mm}g_e(r)Q=\sum_{n}^{\infty}Q^nc^e_{n}(z,\frac{r}{r_0})
\\
&&\hspace{-5mm}g_m(r)zQ=\sum_{n}^{\infty}Q^nc^m_{n}(z,\frac{r}{r_0})
\eea
If the running of coupling constants \eqref{gemEffective} has RGS, then the coefficient $c^{e,m}_{n}[z,r/r_0]$'s should satisfy the following structure equation(derived along the routine of \cite{classicConfinement}, III.A),
\beq{}
\frac{dc^{e,m}_n[z,x]}{dx}=\sum_{k=1}^nkc^{e,m}_k[x]c^{e,m\prime}_{n+1-k}[0]
,~x=\ln\frac{r}{r_0}
\eeq
This is a highly non-trivial constraints on the definition of effective couplings \eqref{gem0definition}-\eqref{gemEffective}. But regretfully, we find that except the lowest two levels of coefficients $c_1^{e,m}[z,x]$, $c_2^{e,m}[z,x]$, all other higher coefficients do not satisfy this RGS equation. Maybe, $J_2$, $K_2$ and $Q$ are not connected through so simple a relation like \eqref{gem0definition}. For example, they maybe connected by the following relation
\begin{subequations}
\bea{}
&&\hspace{-7mm}
\big[(r\!J'\!-\!J)^{\!2}\!+\!2\!J^2\!K^2\!+\!2r^{\!2}\!K^{\!\prime2}\!+\!(K^{\!2}\!-\!1)^2\big]=\alpha(r)Q^2
\\
&&\hspace{-7mm}\alpha(r=r_0)=1,~F[J_2r_0^2,K_2r_0^2]=Q
\eea
\end{subequations}
with $F(x,y)$ being a properly chosen function different from $\sqrt{\alpha(r_0)}$ so that the running of $\alpha(r)$ has exact RGS.

Assuming that such a function exist, then from the basic equation of motion \eqref{eqQuark} and the solution \eqref{confiningSolution1}, we will be able to derive out a classic beta function,
\bea{}
\beta\equiv r\frac{d\alpha}{dr}\rightarrow\Big\{
\begin{array}{l}
4\alpha,~\alpha\rightarrow0
\\
\alpha+\frac{1}{e}{ProductLog}[e\alpha],~\alpha\rightarrow\infty
\end{array}
\label{exBetaFunction}
\eea
where $e=2.71828\cdots$, while $ProductLog[x]$ is defined as the inverse function of $y=x\ln[x]$. In serial form \cite{wolframFunctionSite}
\beq{}
ProductLog[x]=\ln{x}-\ln\Big[\ln{x}\Big]-\frac{\ln\big[\ln{x}\big]}{\ln^2{x}}-\cdots
\nonumber
\eeq
Eq\eqref{exBetaFunction} is both a good and a bad result. It's good because it gives us a non-perturbative beta function at strong coupling. Although classic, it has the same scaling $\beta(\alpha)\propto\alpha$ as those \cite{npQCDBetafunction} inspired by supersymmetric Novikov-Shifman-Vainshtein-Zakharov(NSVZ) instanton analysis and supported by lattice results\cite{latQCDBetafunctionSupport}. It's bad because at small couplings limit, its scaling in $\alpha$ does not match the quantum perturbation results, $\beta(\alpha)_\mathrm{qunt}\xrightarrow{\alpha\rightarrow0}{\beta_0\alpha^2}$. As can be easily proved that, the quantum perturbation $\beta$ requires $\alpha(r)\propto[\ln(r_0/r)]^{-1}$, with $r_0$ denoting some ultraviolet cut off. Maybe a natural question is, can we set $J(r),|K(r)|-1\stackrel{r\rightarrow0}{\propto}[\ln(r_0/r)]^{-\frac{1}{2}}$ and solve the basic equations \eqref{eqQuark} to get a more correct strong coupling beta function? The answer is, we cannot do that way. Because, in such cases $r^2J''_{r\rightarrow0}$ has different signs with $J_{r\rightarrow0}$, so the equation $r^2J''-2JK^2=0$ cannot be kept consistently at the starting point of integration. It seems that as long as asymptotical freedom is required, the classic $\beta\xrightarrow{\alpha\rightarrow0}{4\alpha}$ is an unavoidable conclusion.

\section{Summaries and discussions}
As summaries of this paper, we first make a short introduction to Dvali et als ideal of observing running couplings and confinings at purely classic levels and then make a small try to apply this ideal to Yang-Mills theories coupled with external quarks. We construct three solutions to the classic equation of motion in this theory. The first two are exact but non-confining. While the third one manifests both asymptotical freedom and confinings. It is of serial form and exact at the near-source region, but approximate and analytical at the far-away region. From this solution we derive out a non-perturbative beta function describing the running of effective coupling constant, which is linear in the couplings both in the infrared and ultraviolet region. Our result is sensible to understand confinings of QCD at purely classical levels and non-perturbatively, thus provide us intuitions on QCD confinings by the conventional quantum treatment \cite{masudConfinement}. It may provide new ingredients for studies of QCD using gauge/gravity dualities \cite{AdSQCD}.  More directly, we wish our work be helpful on the road of looking for fully analytical solutions to the Yang-Mills equation\cite{jormaReview}. In another word, we wish our work could play roles as that of Julia and Zee \cite{SU2monopole}, which makes valuable preparations for Prasad and Sommerfield's discoveries \cite{PSsolution}.

As a discussion, we must point out that, from purely numerical aspects, eq\eqref{eqQuark} also allows other solutions which are different from what we provide in this paper. For example, it can be easily verified that if one set $J_{r\rightarrow r_0}=r^{-1}r_0$, $K_{r\rightarrow r_0}=1$, one can also get reasonable numerical results as long as he or she limit itself in the region $r_0<r<\infty$. More numerical solutions are possible if $K_{r\rightarrow0}\rightarrow\!\!\!\!\!\!\!\slash~~1$. We numerically find that all such solutions are confining but not necessarily asymptotic free. This point to a conclusion that, confinings have no direct relation with asymptotical freedom, which just verifies the viewpoint of \cite{topOriginConfinement}. The real deep reason of confinement is the topological structure of the gauge group. 

\section*{Acknowledgements}
This work is supported by Beijing Municipal Natural Science Foundation, Grant. No. Z2006015201001.

\section*{Appendix}
In numeric practices, we find that no matter how we set the $r\rightarrow r_0\rightarrow0$ boundary condition, as long as the equation \eqref{eqQuark} can be smoothly integrated from $r=r_0$ to $r\rightarrow\infty$, we always get confining solutions at $r\rightarrow\infty$.  So it maybe very interesting if we can prove that $J(r)\propto r\ln(r)$analytically, since it means that we prove the confinings of classic Yang-Mills theory somehow.  Let us try the following idea. Consider possibilities,
\bea{}
a)&&K^2-1-J^2\stackrel{r\rightarrow\infty}{\propto}-r^{p}
\label{caseA}
\\
b)&&K^2-1-J^2\stackrel{r\rightarrow\infty}{\propto}+r^{p}
\label{caseB}
\\
c)&&K^2-1-J^2\stackrel{r\rightarrow\infty}{\propto}-r^2\ln^2(r)
\label{caseC}\\
d)&&K^2-1-J^2\stackrel{r\rightarrow\infty}{\propto}+r^2\ln^2(r)
\label{caseD}
\eea
where $p$ is allowed to take any real values. Of course, enumeration of $K^2-1-J^2$'s asymptotical behavior cannot be completely listed. But through careful examinations in the following, we will see that for self-consistencies, further listing of other possibilities such as $e^{pr}$ is of no use.
Substituting these possibilities into equation \eqref{eq2Quark}, we will get 
\bea{}
a_1)&&K\propto C_1r^\frac{1}{2}J_{-\frac{1}{p}}[\frac{2}{p}r^\frac{p}{2}]
+C_2r^\frac{1}{2}J_{\frac{1}{p}}[\frac{2}{p}r^\frac{p}{2}]
\nonumber\\
&&~~~\xrightarrow{r\rightarrow\infty}r^{\frac{1}{2}-\frac{p}{4}}
\eea
\bea{}
b_1)&&K\propto C_1r^\frac{1}{2}I_{-\frac{1}{p}}[\frac{2}{p}r^\frac{p}{2}]
+C_2r^\frac{1}{2}I_{\frac{1}{p}}[\frac{2}{p}r^\frac{p}{2}]
\nonumber\\
&&~~~\xrightarrow{r\rightarrow\infty}r^{\frac{1}{2}-\frac{p}{4}}e^{\frac{2}{p}r^{p/2}}
\\
c_1)&&K\xrightarrow{r\rightarrow\infty}C_1\frac{\cos[r\ln(r)]}{\sqrt{1+\ln(r)}}+C_2\frac{\sin[r\ln(r)]}{\sqrt{1+\ln(r)}}
\\
d_1)&&K\xrightarrow{r\rightarrow\infty}C_1\frac{e^{r\ln(r)}}{\sqrt{1+\ln(r)}}+C_2\frac{e^{-r\ln(r)}}{\sqrt{1+\ln(r)}}
\eea
where $I_\nu(x)$ and $J_\nu(x)$ denote the usual Bessel functions.

In case a), if $p=2$ then both the two equations \eqref{eq1Quark}-\eqref{eq2Quark} can be solved exactly with results $J\propto C_1r^2+C_2r^{-1}$, $K\propto D_1\cos[r]+D_2\sin[r]$. So $K^2-1-J^2\propto-r^4+\mathrm{O}[r]$, which is obviously inconsistent with the assumption \eqref{caseA}. So we need only to explore the fact that
\bea{}
&&\mathrm{if}~p>2,~\begin{array}{l}
\mathrm{eq}\eqref{caseA}\Rightarrow J\xrightarrow{r\rightarrow\infty} r^{p/2},
\\
\mathrm{eq}\eqref{eq1Quark}\Rightarrow J\xrightarrow{r\rightarrow\infty}
r^\frac{p}{4}e^{\frac{4}{2-p}r^{(2-p)/4}}
\end{array}
\label{caseAanalysis1}
\\
&&\mathrm{if}~p<2~~\begin{array}{l}
\mathrm{eq}\eqref{caseA}\Rightarrow J\xrightarrow{r\rightarrow\infty}
\sqrt{r^p+r^{1-p/2}}
\\
\mathrm{eq}\eqref{eq1Quark}\Rightarrow J\xrightarrow{r\rightarrow\infty}
r^\frac{p}{4}e^{\frac{4}{2-p}r^{(2-p)/4}}
\end{array}
\label{caseAanalysis2}
\eea
In either condition, the final conclusions from equations \eqref{caseA} and \eqref{eq1Quark} are inconsistent. So, the a) case can be excluded. Similarly, the case b) can also be excluded. But case c) cannot be excluded by self-consistency. While in case d)
\bea{}
&&\mathrm{if}~C_1=0,~\begin{array}{l}
\mathrm{eq}\eqref{caseD}\Rightarrow J^2\xrightarrow{r\rightarrow\infty}-r^2\ln^2(r),
\\
\mathrm{eq}\eqref{eq1Quark}\Rightarrow J~~~\slash\!\!\!\!\!\!\!\!\!\!\xrightarrow{r\rightarrow\infty}
r\ln r
\end{array}
\\
&&\mathrm{if}~C_2=0,~\begin{array}{l}
\mathrm{eq}\eqref{caseD}\Rightarrow J\xrightarrow{r\rightarrow\infty}\frac{e^{r\ln(r)}}{\sqrt{1+\ln(r)}}<K,
\\
\mathrm{eq}\eqref{eq1Quark}\Rightarrow J~~~\slash\!\!\!\!\!\!\!\!\!\!\xrightarrow{r\rightarrow\infty}
\frac{e^{r\ln(r)}}{\sqrt{1+\ln(r)}}
\end{array}
\eea
So this is also excluded. The only  possibility which cannot be excluded is c). While through observing equations\eqref{caseA}, \eqref{caseAanalysis1} and \eqref{caseAanalysis2}, we know that for self-consistency, the asymptotic of $K^2-1-J^2$ should be something between $-r^2$ and $-r^{2+\epsilon}$, but cannot be of the $r^p$ form. So, up to corrections of order $r^2\ln[r]$, $-r^2\ln^2(r)$ is almost the unique  possibility admitted by self-consistency. This finishes our provement of confinings of Yang-Mills theory at the purely classic levels.


\begin{thebibliography}{99}

 \bibitem{SU2monopole}
 't Hooft,
 ``Magnetic Monopoles in Unified Gauge Theories'',
 {\em Nucl. Phys.} {\bf B79} (1974) 276;
 B. Julia and A. Zee,
``Poles with both magnetic and electric charges in non-Abelian gauge theory''
{\em Phys. Rev.} {\bf D11} (received, Dec. 1974) 2227 ;

 \bibitem{SU3monopole}
 W. J. Marciano and Heinz Pagels,
 ``Classical SU(3) gauge theory and Magneic monopoles'',
 {\em Phys. Rev.} {\bf D12} (1975) 1093;

 Horvath and L. Palla,
 ``Dyons in classical SU(3) gauge theory and a new topologically conserved quantity'',
 {\em Phys. Rev.} {\bf D14} (1976) 1711

 \bibitem{PSsolution}
 M. K. Prasad and C. M. Sommerfield,
 ``Exact classical solution for the 't Hooft Monopole and the Julia-Zee Dyon'',
 {\em Phys. Rev. Lett.} {\bf35} (June, 1975) 760.
 
 \bibitem{sphericalSymmetricYMfield}
 H. P. Kunzle,
 ``SU(n)-Einstein-Yang-Mills fields with spherical symmetry'',
 {\em Class. Quant. Grav.} {\bf8} (1991) 2283;

 J. E. Baxter, Marc Helbling, Elizabeth Winstanley
 ``Soliton and black hole solutions of su(N) Einstein-Yang-Mills theory in anti-de Sitter space'',
 {\em Phys. Rev.} {\bf D74} (2007) 104017
 e-Print: \href{http://arxiv.org/abs/0708.2357v2}{{\tt 0708.2357}}

 \bibitem{particleLikeSolution}
 R. Bartnik and J. Mckinnon,
 ``Particlelike Solutions of the Einstein-Yang-Mills Equations'',
 {\em Phys. Rev. Lett.} {\bf61}  (1988) 141;

 P. Bizon,
 ``Colored Black Holes'',
 {\em Phys. Rev. Lett.} {\bf64} (1990) 2844.

 \bibitem{Yang-Mills-BH}
 D. Singleton, 
 "Exact Schwarzschild - like solution for Yang-Mills theories"
 {\em Phys. Rev.} {\bf D51} (1995) 5911, 
e-Print: \href{}{\tt hep-th/9501052};
"Exact Schwarzschild - like solution for SU(N) gauge theory"
Z.Phys.C72:525,1996,
e-Print: hep-th/9501097;
"General relativistic analog solutions for Yang-Mills theory",
{\em Theor.Math.Phys.} {\bf117} (1998) 1351-1363, 
e-Print: hep-th/9904125 
D. Singleton and A. Yoshida, 
"Increasing potentials in nonAbelian and Abelian gauge theories",
{\em Int. J. Mod. Phys.} {\bf A12} (1997) 4823. 
e-Print: hep-th/9509053.
 
 \bibitem{gaussLawYM}
 V. Lahno, R. Zhdanov and W. Fushchych,
 ``Gauss's Law in Yang-Mills Theory'',
 {\em Academic Dissertation} {University of Helsinki} (2005) Report Series in Physics.
 
 \bibitem{yangMillsCurvedSpace}
 J. A. Sanchez-Monroy, C. J. Quimbay,
 ``Exact solutions of (n+1)-dimensional Yang-Mills equations in Curved space-time'',
 e-Print: \href{http://arxiv.org/abs/1206.3013v1}{{\tt 1206.3013}}.
 
 \bibitem{solByMathematicians1}
 V. Lahno, R. Zhdanov and W. Fushchych,
 ``Symmetry Reduction and Exact Solutions of the Yang-Mills Equations'',
 {\em Nonlinear Mathematical Physics} {V2} (1995) No. 1, 1-23.
 
 \bibitem{solByMathematicians2}
 J. Jormakka,
 ``Solutions to Yang-Mills Equations'',
 e-Print:\href{http://arxiv.org/abs/1011.3962}{{\tt1011.3962}}.
 
 \bibitem{classicConfinement}
 Gia Dvali, Cesar Gomez, Slava Mukhanov,
 ``Classical Dimensional Transmutation and Confinement'',
 e-Print: \href{http://arxiv.org/abs/1107.0870v1}{{\tt 1107.0870 }}.
 
 \bibitem{EFTideal}
W. D. Goldberger and I. Z. Rothstein,
``An Effective field theory of gravity for extended objects'',
{\em Phys. Rev.} {\bf D73} (2006) 104029,
e-Print: hep-th/0409156.

 \bibitem{wolframFunctionSite}
 \href{http://functions.wolfram.com/ElementaryFunctions/ProductLog/}{{\tt http://functions.wolfram.com/ElementaryFunctions}} {\tt/ProductLog/}

 \bibitem{npQCDBetafunction}
 T. A. Ryttov and F. Sannino, 
 ``Supersymmetry inspired QCD beta function'',
 {\em Phys. Rev.} {\bf D78}, 065001(2008).
 M. Frasca, ``Infrared behavior of the running coupling in scalar field theory'',
 e-Print: \href{http://arxiv.org/abs/0802.1183v4}{{0802.1183}}
 \bibitem{latQCDBetafunctionSupport}
 P. Boucaud, F. De Soto, A. Le Yaouanc, J. P. Peroy, J. Micheli, H. Moutarde, O. Pene and J. Rodriguez-Quintero, 
 ``The strong coupling constant at small momentum as an instanton detector'',
 {\em J. High Energy Phys.} {\bf04}(2003)005.
 
 \bibitem{masudConfinement}
 M. Chaichian, K. Nishjima, ``An Essay on Color Confinement'',
 e-Print: \href{http://arxiv.org/abs/hep-th/9909158v1}{{\tt hep-th/9909158}};
 ``The Goto-Imamura-Schwinger Term and Renormalization Group'', 
 {\em Sci. Rev. }{\bf J31} (1999)57, 
 e-Print: \href{http://arxiv.org/abs/hep-th/9909159v1}{{\tt hep-th/9909159}};  
 ``Renormalization Constant of the Color Gauge Field as a Probe of Confinement'',
 {\em Eur. Phys. J.} {\bf C22} (2001) 463,
 e-Print:\href{http://arxiv.org/abs/hep-th/0010079v2}{{\tt hep-th/0010079}}; 
 ``Significance of the Renormalization Constant of the Color Gauge Field'',
 e-Print: \href{http://arxiv.org/abs/hep-th/0302208v1}{{\tt hep-th/0302208}};
K. Nishjima, A. Tureanu,
``Gauge-Dependence of Green's Functions in QCD and QED''
{\em Eur. Phys. J.} {\bf C53} (2008) 649,
e-Print: \href{http://arxiv.org/abs/0710.1257v2}{{\tt0710.1257}}.
 
 \bibitem{AdSQCD}
 U. Gursoy and E. Kiritsis, 
 ``Exploring improved holographic theories for QCD: Part I'',
 {\em J. High. Energy Phys.} {\bf 02} (2008) 032;
 U. Gursoy, E. Kiritsis and F. Nitti, 
 ``Exploring improved holographic theories for QCD: Part II'',
 {\em J. High Energy Phys.} {\bf 02}(2008) 019;
 D. F. Zeng, 
 ``Heavy quark potentials in some renormalization group revised AdS/QCD models'',
 {\em Phys. Rev.} {\bf D78} (2008) 126006.
   
  \bibitem{jormaReview}
 Jorma Jormakka,
 ``Solutions to Yang-Mills equations'',
 e-Print: \href{http://arxiv.org/abs/1011.3962v2}{{\tt 1011.3962 }}.

 \bibitem{topOriginConfinement}
 A. Polyakov, 
 {\em Phys. Letters} {\bf B59} (1975) 82;
 {\em Nucl. Phys. } {\bf B120} (1977) 429.

 \end{thebibliography}
 \end{document}